\documentclass[conference]{IEEEtran}

\usepackage{cite}
\usepackage{amsmath,amssymb}
\usepackage{booktabs}
\usepackage{array}
\usepackage{tabularx}
\usepackage{multirow}
\usepackage{graphicx}
\usepackage[table]{xcolor}
\usepackage{tikz}
\usetikzlibrary{positioning,calc,decorations.pathreplacing,arrows.meta,fit}
\usepackage[most]{tcolorbox}
\usepackage{microtype}
\usepackage[hidelinks]{hyperref}
\renewcommand{\thesection}{\arabic{section}}
\renewcommand{\thesubsection}{\thesection.\arabic{subsection}}
\renewcommand{\thesubsectiondis}{\thesubsection}

\newcommand{\code}[1]{\textsf{#1}}

\newcommand{\motive}[1]{\textit{#1}}

\newtcolorbox{answer}[1]{
  enhanced, breakable=false,
  colback=black!4, colframe=black!45, boxrule=0.5pt, arc=1.5pt,
  left=4pt, right=4pt, top=3pt, bottom=3pt,
  fonttitle=\bfseries\small, coltitle=black, colbacktitle=black!12,
  title={Answer to #1}, fontupper=\small,
}

\newtcolorbox{Observation}[1]{
  enhanced, breakable=false,
  colback=black!4, colframe=black!45, boxrule=0.5pt, arc=1.5pt,
  left=4pt, right=4pt, top=3pt, bottom=3pt,
  fonttitle=\bfseries\small, coltitle=black, colbacktitle=black!12,
  title={Observation #1}, fontupper=\small,
}

\definecolor{directive}{HTML}{2a78d6}
\definecolor{information}{HTML}{eb6834}
\definecolor{scaffolding}{HTML}{1baf7a}
\definecolor{userprompt}{HTML}{898781}

\newcommand{\ctag}[2]{{\setlength{\fboxsep}{1pt}\colorbox{#1!18}{\textsf{\scriptsize\strut #2}}}}

\ExplSyntaxOn
\seq_new:N \l__paper_words_seq
\seq_new:N \l__paper_boxes_seq
\cs_new_protected:Npn \__paper_shade:nn #1#2
  {
    \hbox:n
      {
        \setlength{\fboxsep}{0pt}
        \colorbox{#1}{\rule[-0.29em]{0pt}{1.16em}#2}
      }
  }
\NewDocumentCommand{\seg}{mm}
  {
    \mode_leave_vertical:
    \int_set:Nn \exhyphenpenalty { 0 }
    \int_set:Nn \doublehyphendemerits { 0 }
    \int_set:Nn \finalhyphendemerits { 0 }
    \seq_set_split:Nnn \l__paper_words_seq { ~ } {#2}
    \seq_clear:N \l__paper_boxes_seq
    \seq_map_inline:Nn \l__paper_words_seq
      { \seq_put_right:Nn \l__paper_boxes_seq { \__paper_shade:nn {#1} {##1} } }
    \seq_use:Nn \l__paper_boxes_seq
      { \discretionary{}{}{\__paper_shade:nn {#1} { \c_space_tl }} }
  }
\ExplSyntaxOff

\newif\ifsegmentfigure
\segmentfigurefalse

\newcommand{\nl}{\textcolor{black!50}{\scalebox{0.8}{$\hookleftarrow$}}}



\begin{document}
\title{Hidden in Plain Sight: an Empirical Study about The Instruction Stacks of Coding Agents}


\author{
  \IEEEauthorblockN{Felipe de Sant'Anna Paix\~{a}o, Dhyego Tavares, Eduardo Almeida}
  \IEEEauthorblockA{Federal University of Bahia\\
    \{felipe.paixao, dhyegocruz, eduardo.almeida\}@ufba.br}
  \and
  \IEEEauthorblockN{Iftekhar Ahmed}
  \IEEEauthorblockA{University of California at Irvine\\
    iftekha@uci.edu}
}
\maketitle

\begin{abstract}

Coding agents send their models more than the user's prompt: they add instructions, tool definitions, and environmental context that form an \emph{instruction stack}. This stack guides execution and changes across releases, yet remains largely unexposed to users and insufficiently characterized in research. We present a longitudinal, cross-agent study of the instruction stacks captured in the first model requests of 1,886 releases of Claude Code, Codex, Gemini CLI, Qwen Code, and OpenCode. Under controlled conditions, we analyze their composition and evolution and trace changes in the four open-source agents to documented motivations. Our findings show why examining the system prompt or reporting the model alone provides an incomplete account of agent inputs. Tool definitions dominate three agents' latest requests, while occasional concentrated revisions alter instructions governing tool use, attribution, and coding style. Such revisions can complicate how researchers identify agent-authored contributions and interpret generated code. We also found that only 48\% of examined change units directly explain the motivation for stack revisions. Across agents, shared capabilities emerge despite increasingly distinct wording. These findings expose an evolving, incompletely documented layer of agent design. The findings motivate documenting captured instruction stacks alongside agent releases, models, and user prompts, and preserving the rationale behind their revisions. Our replication package is available at ~\cite{supplementary-materials}.

\end{abstract}

\begin{IEEEkeywords}
coding agents, system prompts, prompt evolution, software evolution, mining software repositories
\end{IEEEkeywords}

\section{Introduction}
\label{sec:introduction}

\begin{figure*}[t]
\centering
\renewcommand{\ctag}[2]{{\setlength{\fboxsep}{0.6pt}\colorbox{#1!18}{\textsf{\footnotesize\strut #2}}}}
\begin{tikzpicture}[
  font=\footnotesize,
  block/.style={draw=black!30, rounded corners=1.5pt, inner xsep=3pt, inner ysep=2pt,
    text width=13.1cm, align=left, anchor=north west},
  brace/.style={decorate, decoration={brace, amplitude=3pt}, semithick, black!55},
  note/.style={font=\footnotesize, align=left, anchor=west, text width=3.4cm},
]
\node[anchor=west, font=\footnotesize] (req) at (0,0)
  {\textbf{Claude Code 2.1.243}\enspace first request to the model, 74{,}078 characters};
\node[block] (sys) at ($(req.south west)+(0,-1.5pt)$) {%
  \texttt{system} field\hfill\textcolor{black!55}{9{,}517 characters}\\[0.5pt]
  \ctag{information}{identity} You are a Claude agent, built on Anthropic's Claude Agent SDK.\\
  \ctag{directive}{engineering guidance} Write code that reads like the surrounding code: match its comment density, \ldots\\
  \ctag{directive}{execution policy} Before deleting or overwriting, look at the target.\\
  \ctag{information}{context facts} \ldots\ Primary working directory: \underline{\texttt{/workspace}}\\
  \ctag{information}{context facts} OS Version: \underline{Linux 7.1.5-arch1-2}\\
  \textcolor{black!55}{\ldots}};
\node[block, draw=none] (umh) at ($(sys.south west)+(0,-2pt)$) {%
  User message\hfill\textcolor{black!55}{327 characters}\\[0.5pt]
  \ctag{directive}{context management} \texttt{<system-reminder>} \ldots\ Today's date is \underline{2026-09-01}. \ldots};
\node[block, draw=black!60, thick, text width=12.85cm] (usr) at ($(umh.south west)+(3pt,0)$)
  {I AM THE USER PROMPT.};
\coordinate (ubot) at ($(usr.south)+(0,-3pt)$);
\node[draw=black!30, rounded corners=1.5pt, inner sep=0pt,
  fit=(umh) (usr) (ubot)] (um) {};
\node[block] (cat) at ($(um.south west)+(0,-2pt)$) {%
  System message\hfill\textcolor{black!55}{5{,}935 characters}\\[0.5pt]
  \ctag{information}{capability catalog} Available agent types for the Agent tool: - claude: Catch-all for any task \ldots\\
  \ctag{information}{capability catalog} - run: Launch and drive this project's app to see a change working.\\
  \textcolor{black!55}{\ldots}};
\node[block, fill=black!5] (tools) at ($(cat.south west)+(0,-2pt)$) {%
  \texttt{tools} field, 20 tool definitions\hfill\textcolor{black!55}{57{,}346 characters}\\[0.5pt]
  \texttt{WebSearch}: ``Search the web. \ldots''\enspace Arguments: \texttt{query}, \texttt{allowed\_domains}, \texttt{blocked\_domains}.\\
  Other tools: \texttt{Agent}, \texttt{Bash}, \texttt{Edit}, \texttt{Read}, \ldots, \texttt{Write}.};
\draw[brace] ($(sys.north east)+(3pt,0)$) -- ($(sys.east |- umh.south)+(3pt,0)$)
  node[note, midway, xshift=6pt] {\textbf{Instruction text}};
\draw[brace] ($(sys.east |- usr.north)+(3pt,0)$) -- ($(sys.east |- usr.south)+(3pt,0)$)
  node[note, midway, xshift=6pt] {\textbf{User prompt}\\0.03\% of the request};
\draw[brace] ($(cat.north east)+(3pt,0)$) -- ($(cat.south east)+(3pt,0)$)
  node[note, midway, xshift=6pt] {\textbf{Instruction text}};
\draw[brace] ($(tools.north east)+(3pt,0)$) -- ($(tools.south east)+(3pt,0)$)
  node[note, midway, xshift=6pt] {\textbf{Tool definitions}};
\end{tikzpicture}
\caption{The first request that Claude Code 2.1.243 sent to its model in our
capture, abridged. The user typed one sentence, which is 0.03\% of the request.
Most of the rest is the instruction stack that the agent adds: instruction
text, in the \texttt{system} field and in messages, and tool definitions.
Colored tags show how our annotation labels the instruction text
(Table~\ref{tab:codebook}). Underlined values come from
the environment where the agent is running: the working directory, the
kernel version of the host, and the date.}
\label{fig:request}
\end{figure*}
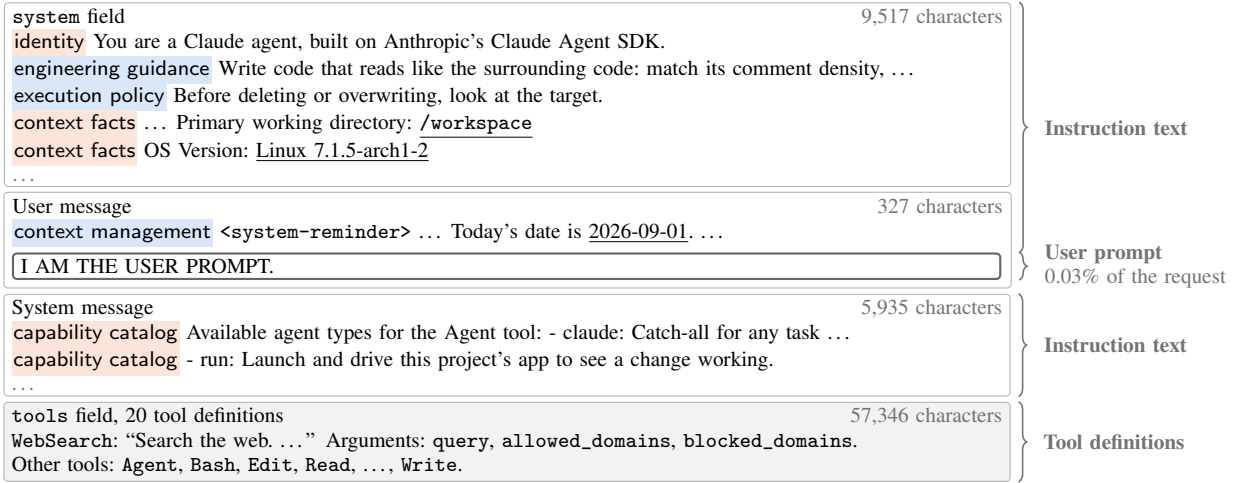

Coding agents such as Claude Code~\cite{anthropicclaudecode}, Codex~\cite{openaicodex}, and Gemini CLI~\cite{googlegeminicli} extend LLM-based software development from generating code suggestions to executing multistep tasks. They inspect repositories, edit files, and run commands on a user's behalf. However, the user's request is only one part of the input that guides this work. Before calling a Large Language Model (LLM), an agent supplements the request with instructions, tool definitions, and context. For example, SWE-agent supplies instructions, documentation, and demonstrations of the commands the model may execute~\cite{yang2024sweagent}. Fig.~\ref{fig:request} illustrates this additional material in Claude Code alongside the user's prompt; the request contains agent-authored instructions, definitions of available tools, and facts collected at runtime, such as the current date and working directory.

We call this agent-supplied material the \emph{instruction stack}. It comprises \emph{instruction text}---behavioral directives, contextual facts, and structural text supplied through top-level fields or messages---and \emph{tool definitions}, including descriptions and argument schemas. The system prompt is therefore only one component of the stack. Our analysis examines the stack in the first request to the main model under a fixed experimental configuration, excluding the user prompt and API configuration parameters. We extracted the instruction stack from widely adopted agents in a fresh, sandboxed environment, allowing us to isolate the agent’s contribution to the stack. This stack guides how the model interacts with users, modifies code, requests permission, and uses tools. Although parts may be available in source code, the assembled stack is generally not exposed in the user interface and can change between releases.

These changes in the \emph{instruction stack} matter because instructions can influence model behavior and resource use. Rewording or reformatting prompts can shift accuracy~\cite{sclar2024formatting,mizrahi2024state}, while increasing instruction count or input length can reduce instruction-following reliability~\cite{jaroslawicz2025instructions,levy2024same,du2025context}. Even changing a tool's description can alter its invocation frequency~\cite{faghih2025tool}. More broadly, a study of 35 Qwen Code releases using a fixed model found that token expenditure per task increased by approximately 70\%~\cite{bensghaier2026harnessevolution}. Although that finding concerns the harness as a whole, it illustrates why model identity alone cannot characterize an agent's behavior or cost.

The \emph{instruction stack} also provides the context in which developer-written guidance operates. Files such as \texttt{AGENTS.md} and \texttt{CLAUDE.md} supply repository-specific instructions~\cite{liu2026diveclaudecode}, and recent studies examine their content and evolution~\cite{chatlatanagulchai2026agentreadmes,mohsenimofidi2026context,jiang2026cursorrules,chakrabarti2026claudemd}. Evidence of their effectiveness is mixed~\cite{gloaguen2026agentsmd,lulla2026agentsmd}, with agent-specific behavior proposed as one explanation~\cite{khatri2026contextfiles}. These files may repeat or contradict vendor instructions, which can themselves conflict~\cite{mason2026arbiter}. Understanding the surrounding stack is therefore necessary to characterize the conditions under which such guidance is evaluated. More generally, researchers who report only the model and user prompt leave an evolving input to agent behavior unspecified.

Prior work provides only a partial view of this input. Prompt-mining studies examine strings and templates embedded in application source code~\cite{pister2024promptset,tafreshipour2025prompting,mao2025templates}, whereas coding agents assemble requests at runtime from multiple instruction sources, tool definitions, and environmental information. Source inspection alone may therefore miss what reaches the model and is unavailable for closed-source agents such as Claude Code. Harness studies describe implementations at one or two points in time~\cite{barbaste2026harness,liu2026diveclaudecode}; the closest longitudinal study evaluates one agent's releases for cost and quality~\cite{bensghaier2026harnessevolution}. To our knowledge, no prior study systematically captures and compares the assembled instruction stacks of multiple production coding agents across their release histories.

These observations motivate a longitudinal, cross-agent analysis. Following release histories reveals which directives persist, change, or disappear and, where development records provide evidence, why maintainers revise them. Comparing agents distinguishes shared practices from agent-specific choices about tool use, verification, autonomy, and user interaction. Together, these analyses characterize the evolution of an often unobserved instruction stack and help characterize the design space in which these agents operate.

We study Claude Code, Codex, Gemini CLI, Qwen Code, and OpenCode. We attempted to run all 1,911 stable npm releases published before 25 August 2026, each in a fresh sandbox with the same user prompt and a fixed configuration per agent. A local server intercepted the first request to the main model for 1,886 releases. We focus on this request because it exposes the initial instruction stack before responses from the main model and subsequent tool results cause execution paths to diverge. Later requests combine instructions with task-dependent histories, making differences harder to attribute to release changes. The first request thus provides a controlled starting point for tracing release-level changes, while excluding instructions loaded later or the full context of interactive sessions. We normalized four API protocols into a common representation, segmented the instruction text into sentences, and classified each segment using a functional codebook. For the four open-source agents, we also traced changes to development records to identify documented motivations. We answer four research questions:

\begin{description}
\item[RQ1.] How are agents' instruction stacks structured?
\item[RQ2.] How do instruction stacks evolve across releases, and what is added or removed?
\item[RQ3.] What documented motivations underlie changes to instruction stacks?
\item[RQ4.] What content is shared across agents, what is agent-specific, and how do these patterns evolve?
\end{description}

This paper makes four contributions:
\begin{enumerate}
\item A capture infrastructure that records initial model requests from isolated agent releases and normalizes four API protocols.
\item A longitudinal dataset of 1,886 captured requests from five production coding agents, with capture provenance.
\item A functional characterization of instruction stacks that distinguishes behavioral directives, contextual information, and structural text.
\item A longitudinal, cross-agent analysis of stack changes, shared practices, and agent-specific choices, including documented motivations for the four open-source agents.
\end{enumerate}

The rest of the paper is structured as follows. Section~\ref{sec:background} defines the terms we use and discusses related work. Section~\ref{sec:methodology} describes how we built the dataset and how we answer each question, and Section~\ref{sec:results} reports the results. Section~\ref{sec:discussion} discusses implications and threats to validity, and Section~\ref{sec:conclusion} concludes.


\section{Background}
\label{sec:background}

A \emph{coding agent} is a program that pairs an LLM with tools for reading, editing, and running code~\cite{yang2024sweagent,wang2025openhands,liu2026agentsurvey,li2025aidev}. It works in a loop of \emph{turns}. In each turn, it sends a request to a model provider, the company that serves the model's API, and the model answers with text or with calls to tools. The agent then runs the tools, appends their results to the conversation, and starts the next turn. Recent work calls this program around the model the agent's \emph{harness}~\cite{bensghaier2026harnessevolution,barbaste2026harness}. Some agents also invoke helper models, for example, to route a task or to write a title for the session. We distinguish these from the \emph{main model}, which handles the user's task, and analyze the first request sent to that model.

Like every other request, this first request is a JSON object whose shape depends on the provider's API protocol. The agents in this study use four protocols: Anthropic Messages~\cite{anthropicmessages}, Gemini Generate Content~\cite{googlegenerate}, OpenAI Responses~\cite{openairesponses}, and OpenAI Chat Completions~\cite{openaichat}. The field names differ, yet the four protocols carry text in the same three kinds of places. First, three have a top-level instruction field, \texttt{system}, \texttt{systemInstruction}, or \texttt{instructions}; Chat Completions has none. The content of this field is commonly called the \emph{system prompt}~\cite{neumann2026controls}. 

Second, all four protocols carry an ordered list of messages, in a field called \texttt{messages}, \texttt{contents}, or \texttt{input}. Each message has a \emph{role} that says whom it speaks for. System and developer messages carry instructions from the application, user messages carry what the user typed, and assistant messages are the model's own earlier replies~\cite{openaitext,anthropicusingmessages}; Gemini calls this last role \texttt{model}. Third, all four protocols carry a list of tool definitions. A \emph{tool definition} gives a tool's name, a description in natural language, and a JSON Schema for its arguments. Some protocols also offer built-in tools that a request declares by type only, such as web search.

Beyond tools, recent agents can extend themselves at run time. A \emph{skill} is a packaged set of instructions that the agent loads when a task needs it~\cite{wang2026malskills}, and \emph{subagents}, separate model sessions to which work can be delegated. Agents announce both to the model as named lists with descriptions, and each agent chooses where to put these lists. For instance, in Fig.~\ref{fig:request}, Claude Code lists its skills in a system message that follows the user prompt. Consequently, an agent's text is spread across several fields and messages, and its layout differs from one agent to another.

We use the following terms to describe these structures consistently, illustrated in the request in Fig.~\ref{fig:request}.

\begin{itemize}
\item The \emph{request} is the complete JSON body that the agent sends to the provider.

\item The \emph{user prompt} is the text supplied by the user. Our captures use the fixed prompt \texttt{I AM THE USER PROMPT}.

\item A \emph{role} says whom a piece of text speaks for. Besides the four message roles, \texttt{system}, \texttt{developer}, \texttt{user}, and \texttt{assistant}, we treat the top-level instruction field as its own role, separate from system messages. The request in Fig.~\ref{fig:request} uses three roles: the top-level instruction, \texttt{user}, and \texttt{system}. Tool definitions have no role, because their text sits inside structured definitions. We analyze them separately and do not segment them.

\item The \emph{instruction stack} comprises \emph{instruction text} and \emph{tool definitions}. The \emph{instruction text} is the text of all roles except the user prompt, including the text that the agent places in user messages. The \emph{tool definitions} are the structured list of tools. Hence, the system prompt is only one part of the instruction text. The remaining fields of a request, such as the model name, generation settings, and JSON syntax, are not part of the stack. Despite its name, the instruction text does more than instruct: it also states facts and marks the structure of the conversation.

\end{itemize}




\section{Related work}
\label{sec:related}

Researchers have studied prompts as software artifacts, examining their content, evolution, and maintenance. PromptSet extracted over 61,000 prompts from open-source Python programs~\cite{pister2024promptset}, while Tafreshipour et al.\ found that only 21.9\% of prompt-changing commits mention the change~\cite{tafreshipour2025prompting}. Other studies classify prompt components~\cite{mao2025templates,basmov2026prompts}, investigate prompt management and technical debt~\cite{li2026promptmanagement,aljohani2025promptdebt}, and propose engineering support for prompts as programs~\cite{liang2025prompts,chen2026promptware,villamizar2025prompts}. These efforts, however, treat prompts as discrete, user-authored artifacts and do not examine the layered, agent-assembled instruction stacks that govern autonomous coding agents at runtime.

As coding agents take on multi-step development tasks, attention has turned to the harnesses that construct model inputs and manage execution. Harness design—tool-result presentation, planning strategies, and context management—demonstrably affects agent performance independently of the underlying model~\cite{yang2024sweagent,wang2025openhands,xia2025agentless,lewis2026harness,fan2026harnessdesign}. Architectural studies characterize harness implementations in the wild~\cite{rombaut2026scaffold,liu2026diveclaudecode}. Barbaste et al.\ compare two snapshots of eleven harnesses and report a shift in behavioral policy from prompt prose to structured configuration~\cite{barbaste2026harness}. Ben Sghaier et al.\ analyze 35 consecutive Qwen Code releases and find rising token expenditure without commensurate improvement in resolve rates~\cite{bensghaier2026harnessevolution}. These studies examine harness architecture and outcomes, whereas we investigate the composition and evolution of the instruction stacks that harnesses send to models.

While the harness implements request construction, tool orchestration, and execution logic, the system prompt provides model-facing behavioral instructions. Research examines these instructions alongside additional guidance supplied by developers. Prompt-extraction studies show that hidden system prompts can often be recovered~\cite{zhang2024effective,hui2024pleak,yu2024assessing,sha2024prompt}, including from coding agents~\cite{xie2025redteaming}. Neumann et al.\ identify seven topics in 1,309 system prompts~\cite{neumann2026controls}, while Mason analyzes conflicting instructions in snapshots from Claude Code, Codex, and Gemini CLI~\cite{mason2026arbiter}. Studies of developer-written guidance examine \texttt{CLAUDE.md} and \texttt{AGENTS.md} files~\cite{chatlatanagulchai2025manifests,chatlatanagulchai2026agentreadmes,mohsenimofidi2026context}, Cursor rules~\cite{jiang2026cursorrules,sun2026cursorrules}, skills~\cite{hong2026skillsmells,gao2026skills}, and other configuration mechanisms~\cite{galster2026configuring}. They report limited attention to security and performance~\cite{chatlatanagulchai2026agentreadmes}, infrequent instruction removal~\cite{chakrabarti2026claudemd}, and mixed evidence of effectiveness~\cite{gloaguen2026agentsmd,lulla2026agentsmd,khatri2026contextfiles}. These studies illuminate system-prompt content and developer customization, but do not trace the surrounding agent-supplied stack across release histories.


Our contribution combines direct request capture with longitudinal analysis of 1,886 releases across five coding agents. Examining instruction text and tool definitions together reveals what reaches the model under our capture conditions, how it changes, and which practices agents share. For the four open-source agents, we also trace revisions to documented motivations. We capture the first request to the main model to compare initial stacks before model responses and subsequent tool results introduce divergent execution histories. This scope provides a common basis for release comparisons while excluding instructions loaded later. The resulting evidence exposes a source of variation that model identifiers and user prompts alone cannot describe, helping researchers account for release-specific directives when constructing datasets and interpreting agent-generated code.


\section{Methodology}
\label{sec:methodology}

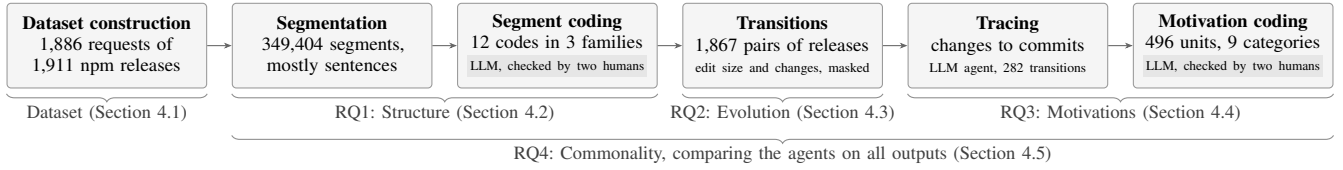
\begin{figure*}[t]
\centering
\begin{tikzpicture}[
  font=\scriptsize,
  box/.style={draw=black!40, fill=black!4, rounded corners=1.5pt, align=center,
    inner sep=2pt, text width=2.5cm, minimum height=1.12cm},
  arr/.style={-{Stealth[length=3pt]}, black!60},
  brace/.style={decorate, decoration={brace, mirror, amplitude=3pt}, black!45},
  rq/.style={font=\scriptsize, text=black!75, inner sep=1pt, below=3pt},
]
\def\checked{{\setlength{\fboxsep}{1pt}\colorbox{black!10}{\tiny\strut LLM, checked by two humans}}}
\node[box] (s1) at (0,0) {\textbf{Dataset construction}\\
  1{,}886 requests of\\1{,}911 npm releases};
\node[box] (s2) at (2.98,0) {\textbf{Segmentation}\\
  349{,}404 segments,\\mostly sentences};
\node[box] (s3) at (5.96,0) {\textbf{Segment coding}\\
  12 codes in 3 families\\\checked};
\node[box] (s4) at (8.94,0) {\textbf{Transitions}\\
  1{,}867 pairs of releases\\{\tiny edit size and changes, masked}};
\node[box] (s5) at (11.92,0) {\textbf{Tracing}\\
  changes to commits\\{\tiny LLM agent, 282 transitions}};
\node[box] (s6) at (14.9,0) {\textbf{Motivation coding}\\
  496 units, 9 categories\\\checked};
\foreach \a/\b in {s1/s2, s2/s3, s3/s4, s4/s5, s5/s6} {\draw[arr] (\a) -- (\b);}
\foreach \a/\b/\name in {s1/s1/Dataset (Section~\ref{sec:method-dataset}),
    s2/s3/RQ1: Structure (Section~\ref{sec:method-rq1}),
    s4/s4/RQ2: Evolution (Section~\ref{sec:method-rq2}),
    s5/s6/RQ3: Motivations (Section~\ref{sec:method-rq3})} {
  \draw[brace] ([yshift=-2pt]\a.south west) -- node[rq] {\name}
    ([yshift=-2pt]\b.south east);
}
\draw[brace] ([yshift=-0.62cm]s2.south west) -- node[rq] {RQ4: Commonality,
  comparing the agents on all outputs (Section~\ref{sec:method-rq4})}
  ([yshift=-0.62cm]s6.south east);
\end{tikzpicture}
\caption{Overview of the methodology. Each step builds on the one before it,
and brackets mark the steps that answer each research question.}
\label{fig:overview}
\end{figure*}

We first capture the request that each release of five agents sends to its model (Section~\ref{sec:method-dataset}). Then, each research question builds on the results of the one before it (Sections~\ref{sec:method-rq1} to~\ref{sec:method-rq4}). Fig.~\ref{fig:overview} gives an overview of the study.

\subsection{Dataset construction}
\label{sec:method-dataset}


\textbf{Agent selection.} We selected widely adopted coding agents that were actively maintained during the study, provided downloadable version histories through a package manager, and supported non-interactive command-line execution. Our sample comprises five agents/harnesses from different vendors: Claude Code (Anthropic)~\cite{anthropicclaudecode}, Codex CLI (OpenAI)~\cite{openaicodex}, Gemini CLI (Google)~\cite{googlegeminicli}, Qwen Code (Alibaba)~\cite{qwencode}, and OpenCode (Anomaly)~\cite{anomalyopencode}. All except Claude Code are open source. These agents also appear in recent coding-agent studies~\cite{vangala2026reproducible,zhang2026pitfalls,bensghaier2026harnessevolution,galster2026configuring,nashid2025multihunk}.



\textbf{Release census.} We enumerated releases from each agent's official npm package. Releases were eligible if published before 25 August 2026, versioned as \texttt{MAJOR.MINOR.PATCH}, and packaged with the agent's executable. This yielded 1,911 eligible releases shown in Table~\ref{tab:dataset}.


\begin{table*}[t]
\centering
\caption{The five agents and their release census. Period: month and year of
the first and last eligible release.}
\label{tab:dataset}
\footnotesize
\begin{tabular}{@{}lllrrl@{}}
\toprule
Agent & npm package & API protocol & Releases & Captures & Period (MM/YY) \\
\midrule
Claude Code & \texttt{@anthropic-ai/claude-code} & Anthropic Messages & 494 & 493 & 02/25--08/26 \\
Codex & \texttt{@openai/codex} & OpenAI Responses & 175 & 174 & 04/25--08/26 \\
Gemini CLI & \texttt{@google/gemini-cli} & Gemini Generate Content & 192 & 192 & 06/25--08/26 \\
OpenCode & \texttt{opencode-ai} & OpenAI Responses & 930 & 907 & 06/25--08/26 \\
Qwen Code & \texttt{@qwen-code/qwen-code} & OpenAI Chat Completions & 120 & 120 & 07/25--08/26 \\
\midrule
Total & & & 1{,}911 & 1{,}886 & \\
 
\bottomrule
\end{tabular}
\end{table*}

\textbf{Controlled capture.} We ran each release once in a fresh \texttt{linux/amd64} container with fixed runtime versions, a new home directory, and an empty working directory. A runner installed the exact release, redirected its provider endpoint to a local HTTP server, and submitted the fixed prompt \texttt{I AM THE USER PROMPT}. This marker identified the user prompt within captured requests. We retained the first request to the main model containing the marker.

Three agents required additional handling because they may first send the prompt to a helper model and, as a result, helper-model requests could also contain the marker. For example, OpenCode and some Claude Code releases ask a model for a session title. Our implementation classified these requests as session-title requests, main-model requests, or unknown. It returned the fixed response title ``Controlled capture session'' and rejected unknown requests. The server stored the selected request's complete JSON body and hashes of both the body and the bytes received over HTTP.

We retained default settings except for endpoint redirection and model selection, where the agent has none by default. Claude Code and Gemini CLI used their default models; Codex used its default \texttt{o4-mini} through release 0.19.0 and \texttt{gpt-5} thereafter. OpenCode used \texttt{gpt-5}, and Qwen Code used \texttt{qwen3-coder-plus}. If a release fails to send its request, we report it in Section~\ref{sec:results}.

\textbf{Request mapping.} To analyze every captured request with one procedure, we map each request to the common terms of Section~\ref{sec:background}. The mapping renames the protocol fields listed there. Thus, \texttt{system}, \texttt{systemInstruction}, and \texttt{instructions} become the top-level instruction, \texttt{messages}, \texttt{contents}, and \texttt{input} become the ordered messages, and each protocol's tool list becomes the tool definitions. Gemini's \texttt{model} role also becomes \texttt{assistant}. We derived these correspondences from each protocol's API reference~\cite {anthropicmessages,googlegenerate,openairesponses,openaichat}. The mapping copies every value unchanged, so it never edits, merges, or reorders text. Moreover, it recognized every role, content type, and tool definition in 1,886 captured requests.






\subsection{RQ1: How are agents’ instruction stacks structured?}
\label{sec:method-rq1}


To answer RQ1, we characterize instruction stacks along three dimensions: \emph{Placement}, the distribution of instruction text across \emph{roles} (Section~\ref{sec:background}); \emph{tools}, the declared tool definitions; and \emph{function}, the purpose of the instruction text. We first count the number of characters inside the requests while grouping by role (e.g., system messages, developer messages). Tools are also an explicit field in the requests these agents make, so we extract tool definitions from their structured fields. However, determining each component's function in the prompt requires reading the text. Hence, we divide the text into segments and code each one following a codebook.



\textbf{Segmentation.} We use sentence-level segments because sentences typically express individual directives or facts. We first divide each role's text into structural blocks, including headings, list items, and paragraphs, then split paragraphs and list items into sentences. Headings and line breaks remain attached to neighboring segments, ensuring that each character belongs to exactly one segment. This procedure yielded 349,404 segments.

\textbf{Codebook.} We built the twelve-code codebook from 25 releases, five per agent. Because agents add and drop parts of their stacks over time, we chose each agent’s first and last releases and the releases at the quartiles of its release order. This purposive sample favors diversity over representativeness~\cite{baltes2022sampling} and is small enough to read in full. We grouped the text of these releases into codes by function, and two authors calibrated the codebook by discussing examples. The codes fall into three families: \emph{directive} text specifies behavior, \emph{information} text states facts, and \emph{scaffolding} text marks conversational structure. Table~\ref{tab:codebook} details the codebook.

To apply the codebook, we drew a random sample (with a fixed seed), stratified by agent and sized for a margin of error of 10 percentage points at 90\% confidence~\cite{baltes2022sampling, cochran1977sampling}. Prior studies that validated automated output against human judgment used the same parameters~\cite{le2024latent,thomas2014studying}. The sample contains 66 requests comprising 12,862 segments. Two authors coded it independently and reached a Cohen’s $\kappa$~\cite{cohen1960coefficient} of 0.75 before adjudication, above our minimum threshold of 0.7~\cite{landis1977measurement}. A third author then resolved their disagreements to produce the reference labels. Our supplementary materials hold the full codebook instructions with 25 worked examples~\cite{supplementary-materials}.


\begin{table}[t]
\centering
\caption{The RQ1 codebook, with twelve codes in three families and
\code{unresolved}, in short. The supplementary materials hold the full definitions.}
\label{tab:codebook}
\footnotesize
\begin{tabularx}{\columnwidth}{@{}l>{\raggedright\arraybackslash}X@{}}
\toprule
Code & The segment states or requires\ldots \\
\midrule

\multicolumn{2}{@{}l}{\textbf{Scaffolding family: structure of the conversation}} \\
\code{user input} & the fixed user prompt, not instruction text \\
\code{scaffolding} & bookkeeping, such as the agent's version \\
\midrule
\addlinespace[2pt]

\multicolumn{2}{@{}l}{\textbf{Directive family: what the model must do, avoid, prefer, or may do}} \\
\code{context management} & how to weigh content in the conversation \\
\code{memory} & how to use memory across sessions \\
\code{execution policy} & permissions and hard-to-reverse operations \\
\code{tool guidance} & when and how to invoke a tool or subagent \\
\code{communication guidance} & what the user reads, and in which form \\
\code{engineering guidance} & how to analyze, change, and verify code \\
\code{role and scope} & which tasks to do, and how autonomously \\
\midrule
\addlinespace[2pt]

\multicolumn{2}{@{}l}{\textbf{Information family: what the model is told}} \\
\code{capability catalog} & named lists of skills, subagents, or tools \\
\code{identity} & the agent's name, product, maker, or model \\
\code{context facts} & facts on the environment or the session \\

\addlinespace[2pt]

\code{unresolved} & a function that the text does not reveal \\

\bottomrule
\end{tabularx}
\end{table}


\textbf{LLM annotation.} We used GLM 5.3 Flash\footnote{Model \texttt{z-ai/glm-5.3-flash}, accessed through OpenRouter with low reasoning effort, the provider's default temperature, and a maximum of 8,192 output tokens.} to assign one code to each segment. Its input included the codebook, the target segment, and up to 15 preceding and following segments for context. A JSON schema constrained responses to one valid code. Before annotating the full dataset, we validated the LLM against the adjudicated reference produced through independent annotation by two human reviewers. The LLM matched the reference on 93.4\% of segments ($\kappa = 0.92$), with every code achieving $\kappa \geq 0.79$. These results met the standard threshold used in practice ~\cite{landis1977measurement}, hence, we use the LLM for annotating the full dataset.

\subsection{RQ2: How do instruction stacks evolve across releases, and what is added or removed?}
\label{sec:method-rq2}


To answer RQ2, we compare consecutive eligible releases in npm publication order. Each pair forms a \emph{transition}, identified by its later release, for example, transition 2.1.69 goes from 2.1.68 to 2.1.69. A transition is \emph{measurable} when both requests were captured, yielding 1,867 measurable transitions out of 1,906. We track changes to segments coded in RQ1 and tool definitions and compare each agent's first and latest release.

\textbf{Edit size.} To measure how much a transition changes the request, we represent each request using its top-level instruction text, messages, and tool definitions, with structured content serialized as canonical JSON. For a transition between strings $A$ and $B$, the \emph{edit size} is their character-level Levenshtein distance $d(A,B)$~\cite{levenshtein1966binary}. The \emph{relative edit size}, $100 \cdot d(A,B)/\max(|A|,|B|)$, normalizes this distance by the longer string's length. Summing edit sizes across measurable transitions yields an agent's cumulative \emph{character edits}.


\textbf{Segment changes.} Following approaches for tracking source-code lines across versions~\cite{zimmermann2006cochanged,reiss2008tracking}, we diff the segments of the two requests, matching segments with the same role and text, ignoring whitespace. A segment that is only present in the later request is \emph{added}, and one only present in the earlier request is \emph{removed}. Yet, this diff would report a moved or edited segment as one removal and one addition, so we pair them in two cases. First, we pair removals and additions with identical text as \emph{moves}, including across roles.  Second, a removed and an added segment at the same place, between the same two matched segments of a role, form one \emph{modified} segment if their relative edit size is at most 50\%, a threshold also used by \texttt{git diff}~\cite{gitdiff}; otherwise, it remains a removal and an addition. We match tools by name, classifying them as added or removed when present in only one request, and modified when their definitions differ.

\textbf{Masking.} An exact comparison of the textual contents of the requests would also count values that change while the instructions do not, namely \emph{run-dependent values}, such as the current date shown in Fig.~\ref{fig:request}, and \emph{release stamps}, which state the agent's version. Hence, before computing changes, we analyzed these repeating patterns from the requests and replaced five such values with placeholders: (1) the current date, (2) the temporary directory of the run, (3) the folder where the agent is installed, (4) the host's kernel version, and (5) Claude Code's release version, which is included in its requests. We then report edit size with and without this \emph{masking}.





\subsection{RQ3: What documented motivations underlie changes to instruction stacks?}
\label{sec:method-rq3}

The changes of RQ2 show what developers edit, but not the motivation behind the change. Since developers often explain an edit in a commit message, we trace each change to the commits that made it and codify the motivations that developers wrote in the commit message. Tracing needs the source history, so RQ3 covers the four open-source agents, with the exception of Claude Code, which is closed-source.

\textbf{Tracing.} We attempt to link each captured release to the commit that its npm package records. Of the four agents' 1,393 captured releases, 1,345  could be linked to a commit, so RQ3 traces only transitions whose two releases are both linked. A change of a transition comes from the commits that only one of its two releases contains. Still, these commits often do not contain the exact changed text in the instruction stack. For example, a commit can edit a single line of code to send a model another prompt file, which replaces the model's instructions without editing either file. Likewise, the text can come from a dependency, and a feature flag or a default setting can decide whether it is sent at all. Line-based techniques such as SZZ~\cite{sliwerski2005changes}, which follow the changed lines back to the commits that last edited them, would thus miss these commits. Instead, we used an LLM-based agent to explore the code that builds the request and, for each change, name the commits that made it. This \emph{tracing agent} is mini-swe-agent~\cite{yang2024sweagent} with Gemini 4 Argon\footnote{Model \texttt{google/gemini-4-argon}, called through OpenRouter, with high reasoning effort and the provider's default temperature and output limit.}. We chose mini-swe-agent for its minimal, model-agnostic design. It has also been widely used in recent studies which found it competitive with vendor harnesses such as Claude Code and Codex~\cite{huang2026deepswe, williams2026sweserve, raghavendra2026sweatlas}. The agent works in an offline sandbox with the two requests, their changes, and the Git history up to both releases, and a program checks that every commit it names is between the two releases.

\textbf{Units.} Developers also explain these commits not only in their messages but also in the pull requests that merged them \cite{pantiuchina2020why, paixao2020behind}. Hence, we group the traced commits by pull request into \emph{units}, and a commit that no pull request merged forms a unit of its own. The \emph{evidence} of a unit is then its commit messages, its pull request's title, description, and review threads, and its linked issues.


\textbf{Initial codebook.} To analyze the reasons behind these \emph{units}, two reviewers built the codebook in Table~\ref{tab:motivation-codebook} from a pilot of 20 of the 282 transitions, drawn at random, five for each of the four agents whose repositories are public. Claude Code was excluded because its repository is private, so its commit history cannot be traced. The pilot served only to build the codebook. As in RQ1, two reviewers then independently coded a separate sample, and a judge adjudicated their disagreements.

For each unit, a coder records two things. First, whether the evidence says why the stack changed: \motive{yes} if it gives a reason for the stack change itself, \motive{indirectly} if it gives a reason only for the code change that the stack change comes with, such as a new feature, and \motive{no} if it gives no reason. Second, the motivations that the evidence states.

\begin{table}[t]
\centering
\caption{The motivation codebook. For each unit, the coder answers whether the
evidence says why the stack changed and records every motivation that it
states, each with a quote from the evidence.}
\label{tab:motivation-codebook}
\footnotesize
\newcommand{\rs}{\addlinespace[3pt]}
\begin{tabularx}{\columnwidth}{@{}>{\raggedright\arraybackslash}p{1.65cm}>{\raggedright\arraybackslash}X@{}}
\toprule
Code & Meaning \\
\midrule

\multicolumn{2}{@{}l}{\textbf{Does the evidence say why the stack changed?}} \\
\motive{yes} & it gives a reason for the stack change itself, such as unclear text, a misbehavior, or the needs of a model \\ \rs
\motive{indirectly} & it gives a reason only for the code change that the stack change comes with, such as a new feature or a refactoring \\ \rs
\motive{no} & it gives no reason \\

\addlinespace[6pt]
\midrule
\multicolumn{2}{@{}l}{\textbf{Motivation behind the change\ldots}} \\

\motive{model behavior} & fixes a misbehavior seen in use or in evals, or steers the model toward the behavior that developers want \\ \rs
\motive{context and cost} & saves tokens, turns, or money, keeps the context or the tool list small, or keeps the prompt cache valid \\ \rs
\motive{safety and security} & protects the user, their system, or their data, such as against prompt injection or destructive actions \\ \rs
\motive{new capability} & supports a new or extended feature, tool, skill, mode, or integration of the agent \\ \rs
\motive{model adaptation} & adapts the request to a specific model or provider, such as a new model or a provider's API \\ \rs
\motive{inconsistency with the code} & makes the text true to how the agent, its tools, or its environment work, as the evidence says that it was false or outdated \\ \rs
\motive{maintenance} & comes with work that keeps the program in shape without a new capability, such as a refactoring, a bug fix outside the model's behavior, or a build change \\ \rs
\motive{text quality} & fixes typos, grammar, or formatting, or makes the text clearer or more consistent, without a misbehavior or a false text \\ \rs
\motive{other} & has a stated reason that fits no category above \\

\bottomrule
\end{tabularx}
\end{table}


\textbf{Coding a unit.} We coded each unit once with Gemini 4 Argon, using the same settings as the tracing agent. Its input lists the unit’s changes, followed by the evidence: commit messages, the pull request’s title and description, linked issues, and review threads. The model answers whether the evidence says why the stack changed, names the first source that states the reason, and quotes every stated motivation. A program then checks each quote against the evidence.



\textbf{Validation.} Before applying the tracing and coding pipeline to the full dataset, we validated both steps on 55 non-pilot transitions drawn at random from the 282 (margin of error of 10 percentage points at 90\% confidence). Two reviewers independently assessed the traced change--unit links and coded the motivations without seeing the LLM’s codes. Of the 798 traced links, 795 were correct (99.6\%), and the agent identified at least one correct unit for all but two sampled changes. For motivation coding, every category reached $\kappa \geq 0.88$ between reviewers and $\kappa \geq 0.82$ between the LLM and the reference, above our minimum threshold of 0.7~\cite{landis1977measurement}. These results support applying the pipeline to the full dataset and reporting every category separately.


\subsection{RQ4: What content is shared across agents, what is agent-specific, and how do these patterns evolve?}
\label{sec:method-rq4}


We compare findings from RQ1 through RQ3 across agents using each agent's latest captured release at each month-end. This aligns observations despite different release frequencies. We track cross-agent variation to assess whether the captured stacks become more similar over time.


\textbf{Shared text.} For each ordered pair of agents $(A,B)$, we calculate the proportion of segments in $A$ whose text also occurs as a segment in $B$. As a sensitivity analysis, we also count near-duplicates. To detect them, we use an edit-similarity threshold of 0.8~\cite{lee2022deduplicating}: a segment of $A$ also counts as shared when its closest segment in $B$ has a relative edit size (Section~\ref{sec:method-rq2}) of at most 20\%. This threshold tolerates light edits, such as a product name replaced in an otherwise copied sentence. Shared text thus measures overlap in wording, complementing the functional comparisons below.


\textbf{Shared tools and functions.} We group tools by function, such as file editing or subagent delegation, to compare capabilities despite different names. We record the first captured appearance of each tool category and code identified in RQ1 in each agent, identifying shared and agent-specific functions and their emergence over time. We also compare the stated motivations identified in RQ3 across the four open-source agents.

\section{Results}
\label{sec:results}

We captured requests from 1,886 of 1,911 eligible releases (Table~\ref{tab:dataset}). The remaining 25 failed before sending a request: 15 OpenCode releases could not initialize the model provider under our configuration, and seven releases across three agents contained broken packages or binaries~\cite{supplementary-materials}.

\begin{figure}[t]
\centering
\includegraphics[width=\columnwidth]{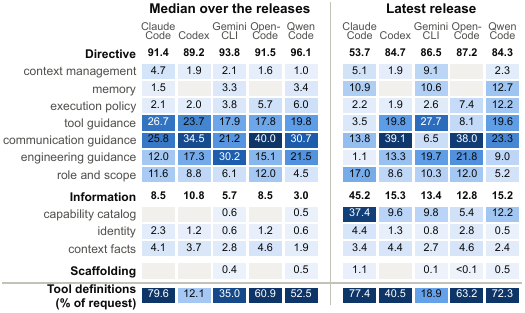}
\caption{Composition of the instruction text (LLM codes), as each code's share of its characters: the median over the agent's captured releases (left) and the share in its latest release (right), on one shading scale. Gray cells contain no code text. Bold rows give the family's share. In the latest release, the codes of a family add up to the family row; the medians are computed row by row and need not add up. Unresolved text is left out, since its median and latest shares round to 0.0\% in every agent. The bottom row, below the horizontal line, is instead the share of the request's characters taken by tool definitions, which are not coded.}
\label{fig:composition}
\end{figure}




\subsection{RQ1: Structure}
\label{sec:results-rq1}

\textbf{Directive text dominates, primarily addressing communication, engineering, and tool use.} Across each agent's releases, the median directive share ranges from 89\% to 96\% of instruction text (Fig.~\ref{fig:composition}, left). Together, \code{communication guidance}, \code{engineering guidance}, and \code{tool guidance} account for median shares of 73\% to 78\%. Communication guidance exceeds engineering guidance in four agents and reaches 40\% in OpenCode. By comparison, \code{execution policy}, covering permissions and risky operations, accounts for 2\% to 6\%. These proportions describe textual allocation, not the importance or effectiveness of the instructions.

\textbf{Tool definitions dominate three agents' latest captured requests.} They take 77.4\% of Claude Code's request, 72.3\% of Qwen Code's, and 63.2\% of OpenCode's, compared with 40.5\% of Codex's and 18.9\% of Gemini CLI's (Fig.~\ref{fig:composition}, bottom row, right). In Claude Code's request, for instance, the 20 tool definitions account for 57,346 of 74,078 characters (Fig.~\ref{fig:request}). Natural-language descriptions constitute most tool-definition content in the first three agents and also convey behavioral instructions. For example, Claude Code's \texttt{Workflow} tool has a 19,290-character description, longer than its entire instruction text, which limits invocation to explicit user requests for multi-agent orchestration. The captured request includes the description despite our prompt not requesting this capability. Overall, Claude Code's tool descriptions are 2.6 times as long as its instruction text, while its system prompt takes only 9,517 characters, 13\% of the request (Fig.~\ref{fig:request}). Hence, system-prompt analysis alone misses substantial instructional content.

\textbf{The user prompt accounts for less than 0.1\% of each agent's latest captured request.} The latest requests (i.e., requests from the latest versions of the agents) range from 29,387 characters in OpenCode to 124,873 in Qwen Code, so our 21-character test prompt (Fig.~\ref{fig:request}) is 0.02\% to 0.07\% of them. Thus, in our experiment, almost all content in these requests is supplied by the agent.

\textbf{Message roles do not reliably identify authorship.} Besides the system prompt, the agent writes messages, each with a role that says whom it speaks for, such as the user or the model (Section~\ref{sec:background}). Although agents place most instruction text in top-level instruction fields or system/developer messages, Gemini CLI and Qwen Code also populate user-role messages. These messages contain 60\% of Gemini CLI's \code{context facts} and 48\% of Qwen Code's; they account for 14\% of Qwen Code's latest instruction text. Both agents also sometimes insert the assistant message ``Got it. Thanks for the context!'', although the model did not generate it. Consequently, a study or tool that reads user messages by looking at messages with the user role might credit the agent's text to the user.

\begin{Observation}{1}
Tool definitions and directive text dominate the captured instruction stacks, and the total stack size can exceed $+120,000$ characters, excluding the user prompt. 
\end{Observation}


\subsection{RQ2: Evolution}
\label{sec:results-rq2}

\begin{figure*}[t]
\centering
\includegraphics[width=0.80\textwidth]{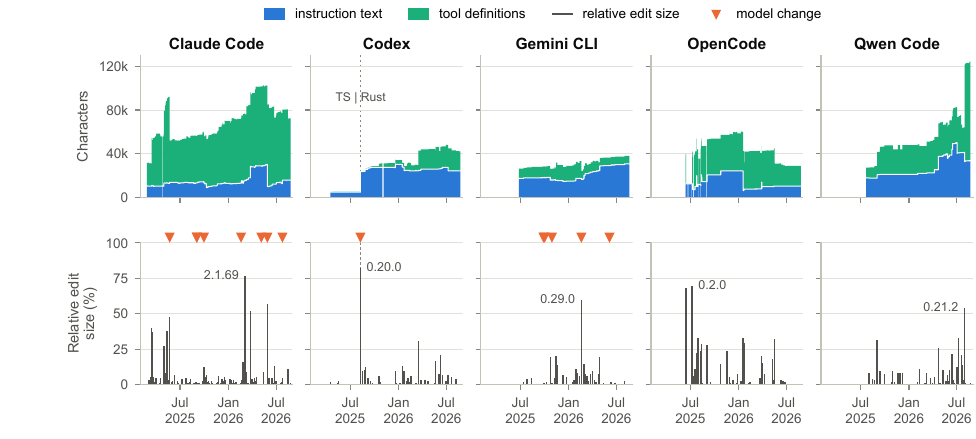}
\caption{Evolution of each agent's request over npm publication time. Top: characters of instruction text and, stacked on it, of tool definitions in each captured release, which together make up at least 93\% of the request; missing captures break the areas. Bottom: relative edit size of each measurable transition, the edit distance between the textual contents of the two requests as a percentage of the longer one. Each bar is one transition, and triangles mark transitions whose request names a different model. Labels name the transition with the largest relative edit size. In Codex, the vertical line marks the switch from the TypeScript to the Rust implementation.}
\label{fig:evolution}
\end{figure*}

\begin{figure}[t]
\centering
\includegraphics[width=0.85\columnwidth]{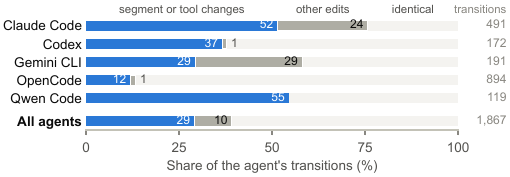}
\caption{Measurable transitions of each agent by what they change. Blue: the transition changes a segment or a tool once run-dependent values and release stamps are masked. Gray: it edits the request but changes no segment or tool. Light gray: the two requests are identical. Labels give the share of the blue and gray parts.}
\label{fig:change}
\end{figure}





\textbf{Request size grew in four agents but shrank in OpenCode.} Between the earliest and latest captured releases, request sizes increased by factors of 7.8 in Codex, 4.5 in Qwen Code, 2.3 in Claude Code, and 1.4 in Gemini CLI; OpenCode's decreased by almost a quarter (Fig.~\ref{fig:evolution}, top). Tool definitions drove the largest component growth in Claude Code and Qwen Code, increasing from 21,005 to 57,346 and from 9,317 to 90,317 characters, respectively.



\textbf{Changes are infrequent and come in bursts.} After masking run-dependent values and release stamps (see Section~\ref{sec:method-rq2}), 544 of 1,867 transitions (29\%) change a segment or tool (Fig.~\ref{fig:change}). Among changing transitions, median relative edit sizes range from 0.6\% to 2.0\% across agents (Fig.~\ref{fig:evolution}, bottom). Only 5 to 10 transitions account for half of each agent's cumulative character edits. These changes mostly add or remove whole segments: the transitions added 3,108 segments and removed 2,751, but modified only 537. As a result, the agents rewrote far more text than their growth shows, adding 306,176 characters of segments and removing 263,685, for a net gain of only 42,491.

\textbf{Large transitions reconfigure tool availability.} Claude Code 2.1.69 replaced 18 tool definitions with a single \texttt{ToolSearch} tool for on-demand discovery, reducing its request from 77,601 to 20,380 characters. Release 2.1.70 restored the definitions a day later. Qwen Code 0.21.2 instead added 65,282 characters of tool definitions, approximately twice the length of its instruction text.

\textbf{Attribution and commenting directives vary across releases.} In 492 of 493 Claude Code captures, the \texttt{Bash} description instructs the model to append a \texttt{Co-Authored-By: Claude} trailer to commit messages, the marker AIDev uses to identify Claude Code pull requests~\cite{li2025aidev}. OpenCode includes a similar directive in 312 of 907 captures, removing it in release 0.4.20 in August 2025. No captured stack from the other three agents specifies an attribution marker. Claude Code's comment policy also changed five times. Its strictest form, ``DO NOT ADD ***ANY*** COMMENTS unless asked'', appears in 186 of 493 captures, spanning releases 0.2.49 to 1.0.128. From release 2.1.154, the policy instead asks the model to ``match its comment density''. Such revisions change the directives governing attribution and coding style without necessarily changing the user's task.

\begin{Observation}{2}
Instruction stacks evolve through occasional concentrated revisions, with additions and removals revealing substantially more change than net size alone captures.
\end{Observation}

\subsection{RQ3: Motivations}
\label{sec:results-rq3}

\begin{figure}[t]
\centering
\includegraphics[width=\columnwidth]{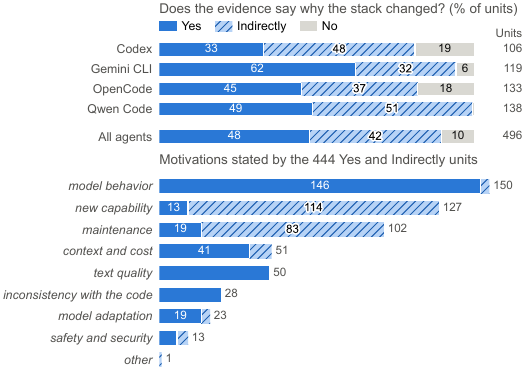}
\caption{Units and their motivations (LLM codes). Top: share of each agent's units by whether their evidence says why the stack changed (Table~\ref{tab:motivation-codebook}). Bottom: number of units that state each motivation, among the 444 units that answer \motive{yes} or \motive{indirectly}, split by that answer. A unit can state several motivations, so the bars add up to 545.}
\label{fig:motivations}
\end{figure}





\textbf{Most units document a reason, but fewer than half explain the stack change itself.} The changes come from 496 units (Section~\ref{sec:method-rq3}), namely 440 pull requests and 56 commits outside any pull request. Of these, 444 (90\%) state a reason (Fig.~\ref{fig:motivations}, top). Yet only 236 (48\%) give a reason for the stack change itself, whereas 208 (42\%) explain only the accompanying code change, such as a new feature. Moreover, units without a pull request explain the least: 21 of the 56 commit-only units (38\%) state no reason, against 31 of the 440 pull-request units (7\%).



\textbf{Stated motivations center on model behavior, capabilities, and maintenance.} Among the 444 units stating a reason, 150 (34\%) cite \motive{model behavior}, 127 (29\%) \motive{new capability}, and 102 (23\%) \motive{maintenance} (Fig.~\ref{fig:motivations}, bottom). Yet these motivations differ in how directly the units explain them. Nearly all units citing \motive{model behavior} (146 of 150) say directly why the stack changed; in fact, they make up 62\% of the 236 units that do, as in Gemini CLI pull request \#17608: ``Fix issue where Gemini CLI can make changes when simply asked a question''. Conversely, most units citing \motive{new capability} (114 of 127) and \motive{maintenance} (83 of 102) explain only the accompanying code change. Hence, developers explain the stack change itself mainly when they steer the model, and otherwise the stack changes along with the surrounding software.

\textbf{Safety is rarely an explicit motivation, including for execution-policy changes.} Only 13 units cite \motive{safety and security}. \code{Execution policy} text, which governs permissions and hard-to-reverse operations, did not shrink between any agent's earliest and latest captured releases and grew by 3,400 characters in Qwen Code. Of the 24 units that modify this text and state a reason, two cite safety, and 13 cite model behavior. Safety is therefore infrequently documented as a motivation, even when changes concern potentially risky operations.



\begin{Observation}{3}
Fewer than half of change units explain the stack revision itself; direct explanations primarily concern model behavior, while safety is rarely an explicit motivation.
\end{Observation}

\subsection{RQ4: Commonality}
\label{sec:results-rq4}

\begin{figure}[t]
\centering
\includegraphics[width=\columnwidth]{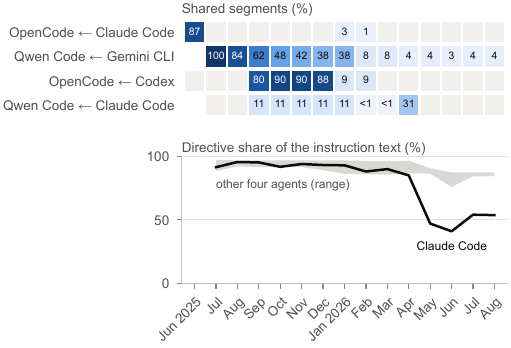}
\caption{Commonality over time. Each column is one month and compares the agents' latest captured releases at its end. Top: for the four pairs of agents that ever share a segment, $A \leftarrow B$ is the share of $A$'s segments that $B$ sends verbatim, where $B$ sent the text first. Segments shorter than 30 characters, the user prompt, and its delimiters are left out. Bottom: share of directive text in the instruction text, across the four agents and Claude Code's line, from July 2025, when all five agents were released.}
\label{fig:commonality}
\end{figure}


\textbf{Shared wording is limited and transient.} Six of ten agent pairs never share an identical segment of at least 30 characters (Fig.~\ref{fig:commonality}, top). Overlap in the remaining pairs reflects adaptation or reuse. Qwen Code originated from Gemini CLI and later incorporated Claude Code's todo-list and memory guidance, accounting for 31\% of Qwen Code's segments in April 2026. In OpenCode, 87\% of segments matched Claude Code in June 2025, while up to 90\% matched Codex between September and December 2025. These overlaps subsequently disappeared as the agents revised their stacks, including changes in the source agents. By August 2026, at least 95\% of each agent's segments appeared in no other agent. Wording therefore became largely agent-specific despite earlier reuse.


\textbf{Capability catalogs and subagent delegation became common across all five agents.} Between the earliest and latest captures, \code{capability catalog} text grew in every agent by up to 5,600 characters. Early catalogs listed user-invoked slash commands such as \texttt{/help}; skill or subagent catalogs appeared between December 2025 (Codex 0.76.0 and Gemini CLI 0.22.0) and June 2026 (Qwen Code 0.18.0). In Claude Code, these catalogs account for 37\% of the latest instruction text. Subagent delegation appeared in Claude Code's first capture and in the other four agents between June 2025 and March 2026. These capabilities became shared features despite limited overlap in wording.

\textbf{Agents share functional categories, while directive shares converge in four agents.} Eight of the eleven RQ1 codes (Table~\ref{tab:codebook}) appear in every agent's latest stack, generally from early releases onward. Agents differ in their allocation of text across these functions and their inclusion of less-common categories (Fig.~\ref{fig:composition}).
The directive shares of the four open-source agents narrowed from 88\% to 97\% in July 2025 to 84\% to 87\% in August 2026 (Fig.~\ref{fig:commonality}, bottom). Claude Code diverged: its directive share fell from 85\% in April 2026 to 54\% following release 2.1.154 (Section~\ref{sec:results-rq2}). Thus, four agents became more similar in directive share despite increasingly distinct wording, while Claude Code moved away from this pattern (Fig.~\ref{fig:commonality}).

\begin{Observation}{4}
Agents share core functions and capabilities despite using increasingly distinct wording.
\end{Observation}

\section{Discussion}
\label{sec:discussion}

\textbf{Instruction stacks need maintenance rationale, not just version histories.} Substantial additions and removals reveal ongoing revision, yet only 48\% of change units explain the change itself. Without this rationale, maintainers may struggle to distinguish enduring requirements from workarounds for earlier models or failures. Retaining obsolete instructions or removing necessary ones could accumulate instruction-related technical debt. Work treating prompts as programs~\cite{liang2025prompts,chen2026promptware} could address this risk by linking revisions to their intended behavior, applicable models, and regression evaluations.

\textbf{Shared practices suggest emerging engineering conventions.} Capability catalogs and subagent delegation became common across all five agents despite increasingly distinct wording. These patterns suggest recurring approaches to exposing capabilities and coordinating work, rather than convergence on a common prompt. They offer a basis for reusable tooling and comparable evaluations, although shared functions alone establish neither interoperability nor effectiveness.

\textbf{The agent release should be part of the experimental configuration.} A coding agent interprets the user's task alongside substantial agent-supplied content, including tool descriptions and messages beyond the system prompt. Because this content changes across releases, model identity and the user prompt alone do not fully specify experimental conditions. Differences attributed to model capabilities may also reflect changes in harness-supplied instructions. Our study exposes this source of variation without establishing its causal effects on performance or cost.

Reporting recommendations for LLM-based studies~\cite{sallou2024breaking,baltes2025guidelines,korn2026reporting} should therefore extend to the exact agent release, model configuration, repository guidance, and enabled capabilities. Where feasible, researchers should also archive captured instruction stacks with their capture conditions. Release-level information matters for repository mining as well: attribution directives can affect which contributions marker-based searches retrieve~\cite{li2025aidev}, while commenting policies offer an alternative explanation for differences in generated code. Accounting for agent releases is thus necessary even when the researcher-written prompt and selected model remain unchanged.

\section{Threats to validity}
\label{sec:threats}


\textbf{Construct validity.} RQ3 captures developers' stated motivations, which may omit or misrepresent their actual reasons~\cite{cai2026rule,paixao2020behind}. We examine linked pull requests, reviews, and issues alongside commit messages and do not infer motivations from code changes alone. Our findings therefore describe stated motivations, not verified intent.

\textbf{Internal validity.} Human judgment and LLM-assisted annotation may introduce labeling errors. We followed established evaluation practices, including independent labeling, adjudication, and validation against human references using predefined acceptance criteria~\cite{ralph2021empirical,ahmed2025can}. We also restricted commit tracing to eligible candidates and checked supporting quotations. These steps reduce annotation errors but cannot eliminate shared biases or guarantee accuracy beyond the validation samples.

\textbf{External validity.} We study five command-line agents under a controlled configuration, capturing only the first request to the main model in an empty working directory without user interaction. This supports release-level comparison but does not capture variation from user customization, alternative model choices, or later interactions. Other agents and configurations may produce different stacks, and shared content across the five agents does not establish broader generality. Our findings cover captured releases published before 25 August 2026 and may not extend to subsequent releases.

\section{Conclusion}
\label{sec:conclusion}


We analyzed the initial instruction stacks of 1,886 releases across five coding agents. The study shows that agent releases revise model-facing instructions beyond the system prompt, often without documenting the reasons, while shared capabilities emerge despite increasingly distinct wording. These changes introduce a source of variation that model identifiers and user prompts alone cannot capture. Researchers should therefore document the agent release and captured stack when reporting experiments, and maintainers should preserve the rationale for instruction changes.

Our analysis characterizes these changes; however, establishing their effects requires controlled experiments. Future work should vary captured stacks while holding models, tasks, and execution conditions fixed to measure changes in correctness, tool use, latency, and cost. Ablations could test whether particular instructions improve outcomes or introduce redundancy and conflicts. Capturing later turns would extend this analysis to repository guidance, dynamically loaded capabilities, and execution history. Our replication package is available at ~\cite{supplementary-materials}.

\bibliographystyle{IEEEtran}
\bibliography{references}
\end{document}